\documentclass[aps,prl,twocolumn,groupedaddress]{revtex4}
\usepackage{graphicx}
\begin{document}

\title{The phonon theory of liquid thermodynamics}
\author{D. Bolmatov$^{1}$}
\author{V. V. Brazhkin$^{2}$}
\author{K. Trachenko$^{1,3}$}
\address{$^1$ School of Physics and Astronomy, Queen Mary University of London, Mile End Road, London, E1 4NS, UK}
\address{$^2$ Institute for High Pressure Physics, RAS, 142190, Troitsk, Moscow Region, Russia}
\address{$^3$ South East Physics Network}

\begin{abstract}
Heat capacity of matter is considered to be its most important property because it holds information about system's degrees of freedom as well as the regime in which the system operates, classical or quantum. Heat capacity is well understood in gases and solids but not in the third state of matter, liquids, and is not discussed in physics textbooks as a result. The perceived difficulty is that interactions in a liquid are both strong and system-specific, implying that the energy strongly depends on the liquid type and that, therefore, liquid energy can not be calculated in general form. Here, we develop a phonon theory of liquids where this problem is avoided. The theory covers both classical and quantum regimes. We demonstrate good agreement of calculated and experimental heat capacity of 21 liquids, including noble, metallic, molecular and hydrogen-bonded network liquids in a wide range of temperature and pressure.

\end{abstract}

\pacs{65.20.Jk, 61.25.Em, 62.50.-p, 05.70.-a}

\maketitle

Research into liquids has a long history starting from the same time when the theory of gases was developed, forming the basis for our current understanding of the gas state of matter \cite{max}. Yet no theory of liquid heat capacity currently exists, contrary to gases or, for that matter, solids, and is not discussed in physics textbooks and student lectures \cite{granato}.

Liquids defy common approaches used to discuss the other two states of matter, gases and solids. Interactions in a liquid are strong, therefore treating them as a small perturbation as is done in the theory of gases is not an option \cite{landau}. Atomic displacements in a liquid are large, therefore expanding the energy in terms of squared atomic displacements and considering the remaining terms small, as is done in the theory of solids, does not appear justified either. Strong interactions in a liquid appear to result in liquid energy being strongly dependent on the type of interactions, leading to a statement that liquid energy can not be calculated in general form \cite{landau}, in contrast to solids and gases where $E=3NT$ ($k_{\rm B}=1$) and $E=\frac{3}{2}NT$ at high temperature.

Previous research into liquids presents an interesting case when long-lived absence of crucial experimental data resulted in using our human experience and intuition about liquids instead. Liquids have been viewed to occupy a state intermediate between gases and solids. Liquids flow, and share this fundamental property with gases rather than solids. As a result, liquids were mostly viewed as interacting gases, forming the basis for previous theoretical approaches \cite{landau,march}. On the other hand, Frenkel noted that the density of liquids is only slightly different from that of solids, but is vastly different from gas densities \cite{frenkel}. Frenkel has subsequently made another important proposition which has further put liquids closer to solids in terms of their physical properties. He introduced liquid relaxation time, $\tau$, as the average time between two consecutive atomic jumps at one point in space, and immediately predicted the solid-like ability of liquids to support shear waves, with the only difference that a liquid does not support all shear waves as a solid does, but only those with frequency larger than $\frac{1}{\tau}$ because the liquid structure remains unaltered and solid-like during times shorter than $\tau$. Frenkel's ideas did not find due support at his time, partly because of the then existing dominating school of thought advocating different views \cite{abrikos}, and consequently were not theoretically developed. It took many years to verify Frenkel's prediction \cite{copley,grim,pilgrim,burkel,rec-review}. With the aid of powerful synchrotron radiation sources that became available fairly recently, and about 50 years after Frenkel's work, it has become apparent that even low-viscous liquids maintain high-frequency modes with wavelengths close to interatomic separations \cite{rec-review}. We note that being eigenstates of topologically disordered liquids, these modes are phonon-like in a sense that they are not entirely harmonic vibrations as in crystals.

The data from the experimental advances above, together with theoretical understanding of the phonon states in a liquid due to Frenkel raise an important question of whether the phonon theory of liquid thermodynamics could be constructed, similar to the phonon theory of solids. Below we develop a phonon theory of thermodynamics of liquids that covers both classical and quantum regimes. We find good agreement between calculated and experimental heat capacity for many liquids in a wide temperature and pressure range.

\section{Results}

\subsection{Theory}

There are two types of atomic motion in a liquid: phonon motion that consists of one longitudinal mode and two transverse modes with frequency $\omega>\omega_{\rm F}$, where $\omega_{\rm F}=\frac{1}{\tau}$ is Frenkel frequency, and the diffusional motion due to an atom jumping between two equilibrium positions. In turn, the phonon and diffusional motion consists of kinetic and potential parts, giving the liquid energy as

\begin{equation}
E=K_l+P_l+K_s(\omega>\omega_{\rm F})+P_s(\omega>\omega_{\rm F})+K_d+P_d
\label{en0}
\end{equation}

\noindent where $K_l$ and $P_l$ are kinetic and potential components of the longitudinal phonon energy, $K_s(\omega>\omega_{\rm F})$ and $P_s(\omega>\omega_{\rm F})$ are kinetic and potential components of the energy of shear phonons with frequency $\omega>\omega_{\rm F}$, and $K_d$ and $P_d$ are kinetic and potential energy of diffusing atoms. The energy of longitudinal mode is the same as in a solid, albeit different dissipation laws apply at low and high frequency \cite{frenkel}. In the next paragraph, we re-write $E$ in the form convenient for subsequent calculations.

$P_d$ can be omitted because $P_d\ll P_s(\omega>\omega_{\rm F})$. This is because the absence of shear modes with frequency $\omega<\omega_{\rm F}$ due to the dynamic Frenkel mechanism implies smallness of low-frequency restoring forces and therefore smallness of low-frequency potential energy of shear modes: $P_s(\omega<\omega_{\rm F})\ll P_s(\omega>\omega_{\rm F})$, where $P_s(\omega<\omega_{\rm F})$ is the potential energy of low-frequency shear modes \cite{heat1,heat2}. Instead of low-frequency shear vibrations with potential energy $P_s(\omega<\omega_{\rm F})$, atoms in a liquid ``slip'' and undergo diffusing motions with frequency $\frac{1}{\tau}$ and associated potential energy $P_d$. $P_d$ is due to the same interatomic forces that $P_s(\omega<\omega_{\rm F})$, giving $P_d\approx P_s(\omega<\omega_{\rm F})$. Combining this with $P_s(\omega<\omega_{\rm F})\ll P_s(\omega>\omega_{\rm F})$, $P_d\ll P_s(\omega>\omega_{\rm F})$ follows. Re-phrasing this, were $P_d$ large and comparable to $P_s(\omega>\omega_{\rm F})$, strong restoring forces at low frequency would result, and lead to the existence of low-frequency vibrations instead of diffusion. We also note that because $P_l\approx P_s$, $P_d\ll P_s(\omega>\omega_{\rm F})$ gives $P_d\ll P_l$, further implying that $P_d$ can be omitted in Eq. (\ref{en0}). Then, $E=K+P_l+P_s(\omega>\omega_{\rm F})$, where total kinetic energy in a liquid, $K$ ($K=\frac{3}{2}Nk_{\rm B}T$ in the classical case), includes both vibrational and diffusional components. $E$ can be re-written using the virial theorem $P_l=\frac{E_l}{2}$ and $P_s(\omega>\omega_{\rm F})=\frac{E_s(\omega>\omega_{\rm F})}{2}$ (here, $P$ and $E$ refer to their average values) and by additionally noting that $K$ is equal to kinetic energy of a solid and can therefore be written, using the virial theorem, as the sum of kinetic terms related to longitudinal and shear waves: $K=\frac{E_l}{2}+\frac{E_s}{2}$, giving $E=E_l+\frac{E_s}{2}+\frac{E_s(\omega>\omega_{\rm F})}{2}$. Finally noting that $E_s=E_s(\omega<\omega_{\rm F})+E_s(\omega>\omega_{\rm F})$, where the two terms refer to their solid-state values, liquid energy becomes

\begin{equation}
E=E_l+E_s(\omega>\omega_{\rm F})+\frac{E_s(\omega<\omega_{\rm F})}{2}
\label{en1}
\end{equation}

Each term in Eq. (\ref{en1}) can be calculated using the phonon free energy, $F_{ph}=E_0+T\sum\limits_i\ln\left(1-\exp\left(-\frac{\hbar\omega_i}{T}\right)\right)$, where $E_0$ is the energy of zero-point vibrations \cite{landau}. In calculating the energy, $E_{ph}=F_{ph}-T\frac{d F_{ph}}{d T}$, we take into account the effect of thermal expansion, important in liquids. This implies $\frac{d\omega_i}{d T}\ne 0$, contrary to the harmonic case, and gives

\begin{equation}
E_{ph}=E_0+\hbar\sum\limits_i\frac{\omega_i-T\frac{d\omega_i}{d T}}{\exp\left(\frac{{\hbar\omega_i}}{T}\right)-1}
\label{en2}
\end{equation}

Using quasi-harmonic approximation Gr\"{u}neisen approximation gives $\frac{{\rm d}\omega}{{\rm d}T}=-\frac{\alpha\omega}{2}$, where $\alpha$ is the coefficient of thermal expansion \cite{tg}. Putting it in Eq. (\ref{en2}) gives

\begin{equation}
E_{ph}=E_0+\left(1+\frac{\alpha T}{2}\right)\sum\limits_i\frac{\hbar\omega_i}{\exp\left(\frac{{\hbar\omega_i}}{T}\right)-1}
\label{en3}
\end{equation}

The energy of one longitudinal mode, the first term in Eq. (\ref{en1}), can be calculated by substituting the sum in Eq. (\ref{en3}), $\sum$, with Debye vibrational density of states for longitudinal phonons, $g(\omega)=\frac{3N}{\omega_{\rm D}^3}\omega^2$, where $\omega_{\rm D}$ is Debye frequency. The normalization of $g(\omega)$ reflects the fact that the number of longitudinal modes is $N$. Integrating from 0 to $\omega_{\rm D}$ gives $\sum=NTD\left(\frac{\hbar\omega_{\rm D}}{T}\right)$, where $D(x)=\frac{3}{x^3}\int\limits_0^x\frac{z^3{\rm d}z}{\exp(z)-1}$ is Debye function \cite{landau}. The energy of two shear modes with frequency $\omega>\omega_{\rm F}$, the second term in Eq. (\ref{en1}), can be similarly calculated by substituting $\sum$ with density of states $g(\omega)=\frac{6N}{\omega_{\rm D}^3}\omega^2$, where the normalization accounts for the number of shear modes of $2N$. Integrating from $\omega_{\rm F}$ to $\omega_{\rm D}$ gives $\sum=2NTD\left(\frac{\hbar\omega_{\rm D}}{T}\right)-2NT\left(\frac{\omega_{\rm F}}{\omega_{\rm D}}\right)^3D\left(\frac{\hbar\omega_{\rm F}}{T}\right)$. $E_s(\omega<\omega_{\rm F})$ in the last term in Eq. (\ref{en1}) is obtained by integrating $\sum$ from 0 to $\omega_{\rm F}$ with the same density of states, giving $\sum=2NT\left(\frac{\omega_{\rm F}}{\omega_{\rm D}}\right)^3D\left(\frac{\hbar\omega_{\rm F}}{T}\right)$. Putting all terms in Eq. (\ref{en3}) and then Eq. (\ref{en1}) gives finally the liquid energy

\begin{equation}
E=NT\left(1+\frac{\alpha T}{2}\right)\left(3D\left(\frac{\hbar\omega_{\rm D}}{T}\right)-\left(\frac{\omega_{\rm F}}{\omega_{\rm D}}\right)^3D\left(\frac{\hbar\omega_{\rm F}}{T}\right)\right)
\label{enf}
\end{equation}
\noindent where we have omitted $E_0$ that includes zero-point vibration energies given by Eq. (\ref{en1}) because below we calculate the derivative of $E$. We note that in general, $E_0$ is temperature-dependent because it is a function of $\omega_{\rm F}$ and $\tau$ (see Eq. (\ref{en1})). However, this becomes important at temperatures of several K only, whereas below we deal with significantly higher temperatures.

Eq. (\ref{enf}) spans both classical and quantum regimes, an important feature for describing the behavior of liquids discussed below.

We note that the presented phonon theory of liquids operates at the same level of approximation as Debye theory of solids, particularly relevant for disordered isotropic systems \cite{landau} such as glasses and liquids. The result for a harmonic solid follows from Eq. (\ref{enf}) when $\omega_{\rm F}=0$, corresponding to infinite relaxation time, and $\alpha=0$. 

\subsection{Comparison with experimental data}

We now compare Eq. (\ref{enf}) to experimental data of heat capacity per atom, $c_v=\frac{1}{N}\frac{d E}{d t}$. We have used the National Institute of Standards and Technology (NIST) database \cite{nist} that contains $c_v$ for many liquids, and have chosen monatomic nobel liquids, molecular liquids, as well as a hydrogen-bonded network liquids. We aimed to check our theoretical predictions in a wide range of temperature, and therefore selected the data at pressures exceeding the critical pressures of the above systems where they exist in a liquid form in the broad temperature range. As a result, the temperature range in which we calculate $c_v$ is about 100--700 K for various liquids (see Figures 1--4). Experimental $c_v$ of metallic liquids were taken from Refs. \cite{metal1,metal2} at ambient pressure ($c_v$ of some of these were calculated previously in Ref. \cite{heat1} on the basis of classical approach only). The total number of liquids considered is 21.

\begin{figure*}
\begin{center}
{\scalebox{0.75}{\includegraphics{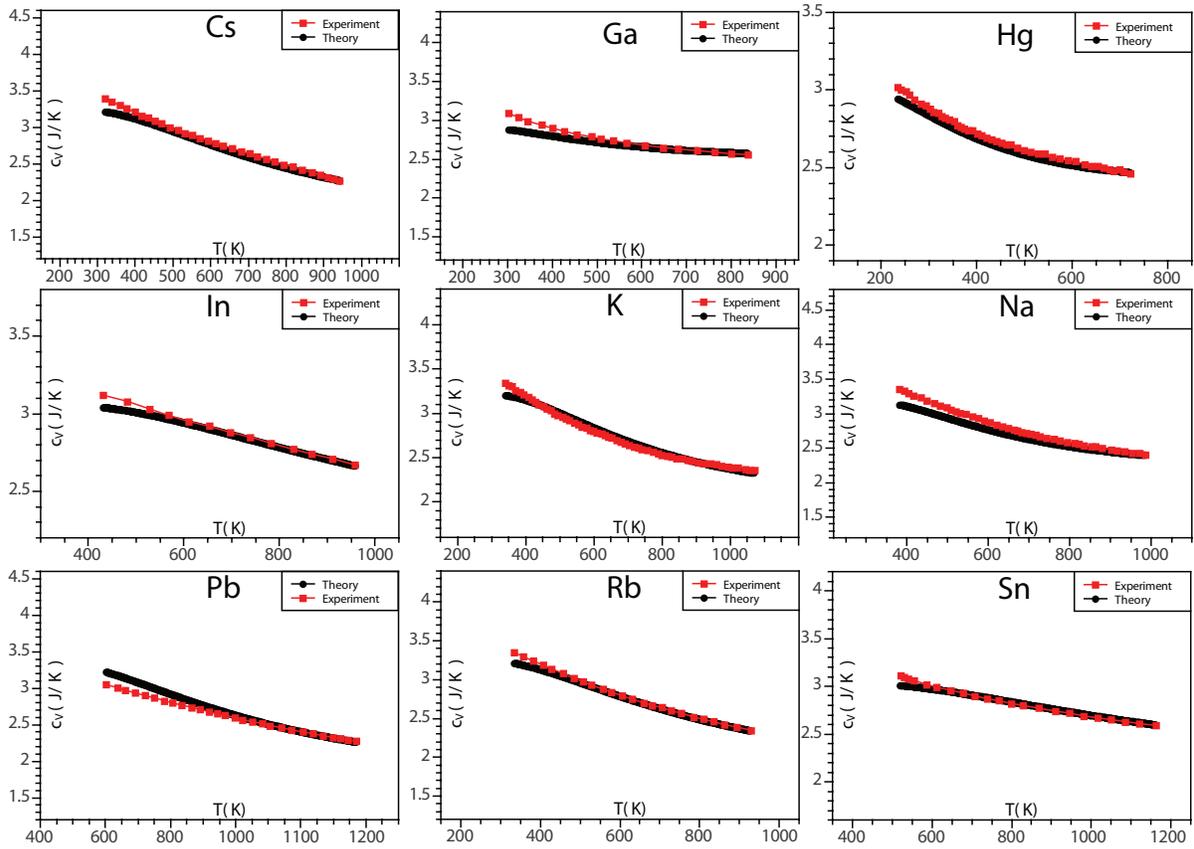}}}
\end{center}
\caption{Experimental and calculated $c_v$ for liquid metals. Experimental $c_v$ are at ambient pressure, with electronic contribution subtracted \cite{metal1,metal2}. Values of $\tau_{\rm D}$ used in the calculation are 0.95 ps (Cs), 0.27 ps (Ga), 0.49 ps (Hg), 0.31 ps (In), 0.57 ps (K), 0.42 ps (Na), 0.64 ps (Pb), 0.74 ps (Rb) and 0.18 ps (Sn). Values of $G_{\infty}$ are 0.17 GPa (Cs), 2.39 GPa (Ga), 1.31 GPa (Hg), 1.58 GPa (In), 1.8 GPa (K), 3.6 GPa (Na), 1.42 GPa (Pb), 0.25 GPa (Rb) and 3 GPa (Sn). Experimental $\alpha$ (Ref. \cite{metal2}) are $3\cdot 10^{-4}$ K$^{-1}$ (Cs), $1.2\cdot 10^{-4}$ K$^{-1}$ (Ga), $1.8\cdot 10^{-4}$ K$^{-1}$ (Hg), $1.11\cdot 10^{-4}$ K$^{-1}$ (In), $2.9\cdot 10^{-4}$ K$^{-1}$ (K), $2.57\cdot 10^{-4}$ K$^{-1}$ (Na), $3\cdot 10^{-4}$ K$^{-1}$ (Pb), $3\cdot 10^{-4}$ K$^{-1}$ (Rb) and $0.87\cdot 10^{-4}$ K$^{-1}$ (Sn). Values of $\alpha$ used in the calculation are $3.8\cdot 10^{-4}$ K$^{-1}$ (Cs), $1.2\cdot 10^{-4}$ K$^{-1}$ (Ga), $1.6\cdot 10^{-4}$ K$^{-1}$ (Hg), $1.25\cdot 10^{-4}$ K$^{-1}$ (In), $2.9\cdot 10^{-4}$ K$^{-1}$ (K), $2.57\cdot 10^{-4}$ K$^{-1}$ (Na), $3\cdot 10^{-4}$ K$^{-1}$ (Pb), $4.5\cdot 10^{-4}$ K$^{-1}$ (Rb) and $1.11\cdot 10^{-4}$ K$^{-1}$ (Sn).}
\end{figure*}

\begin{figure*}
\begin{center}
{\scalebox{0.65}{\includegraphics{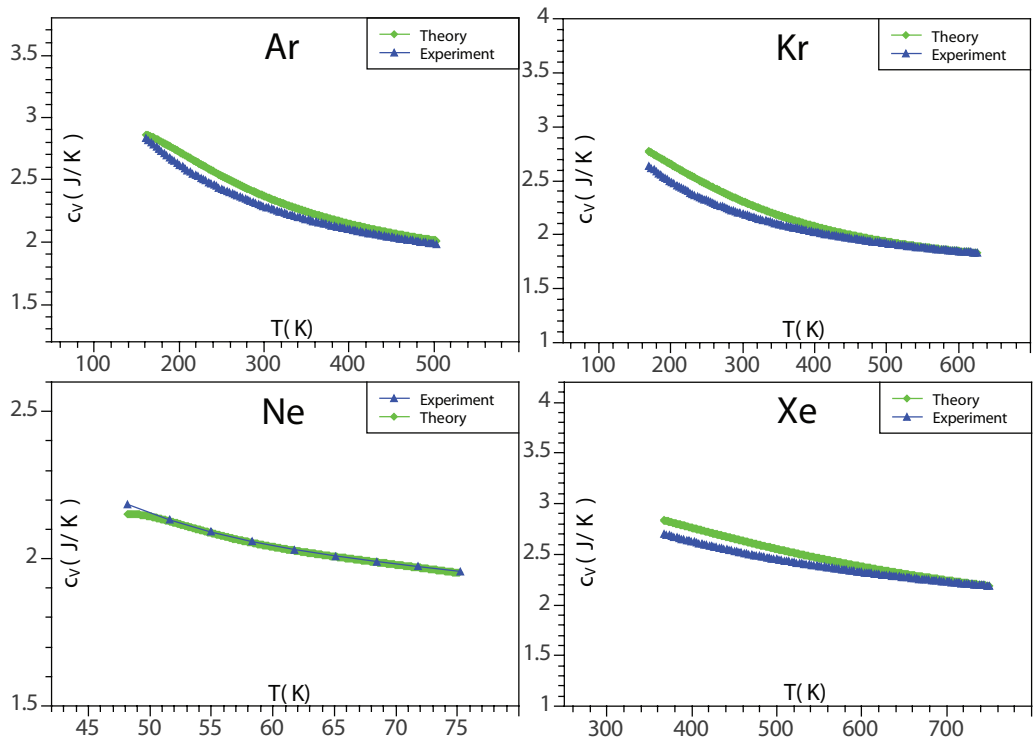}}}
\end{center}
\caption{Experimental and calculated $c_v$ for noble liquids. Experimental $c_v$ and $\eta$ are taken from the NIST database \cite{nist} at pressures 378 MPa (Ar), 200 MPa (Kr), 70 MPa (Ne), and 700 MPa (Xe). Values of $\tau_{\rm D}$ used in the calculation are 0.5 ps (Ar), 0.67 ps (Kr), 0.31 ps (Ne) and 0.76 ps (Xe). Values of $G_{\infty}$ are 0.18 GPa (Ar), 0.23 GPa (Kr), 0.12 GPa (Ne) and 0.45 GPa (Xe). Experimental values of $\alpha$ calculated from the NIST database at the corresponding pressures above are $3.6\cdot 10^{-4}$ K$^{-1}$ (Kr), $7.7\cdot 10^{-3}$ K$^{-1}$ (Ne) and $4.1\cdot 10^{-4}$ K$^{-1}$ (Xe). Values of $\alpha$ used in the calculation are $1.3\cdot 10^{-3}$ K$^{-1}$ (Kr), $6.8\cdot 10^{-3}$ K$^{-1}$ (Ne) and $3.6\cdot 10^{-4}$ K$^{-1}$ (Xe). For Ar, calculating $c_v$ in harmonic approximation ($\alpha=0$) gives good fit to experimental $c_v$.}
\end{figure*}

\begin{figure*}
\begin{center}
{\scalebox{0.65}{\includegraphics{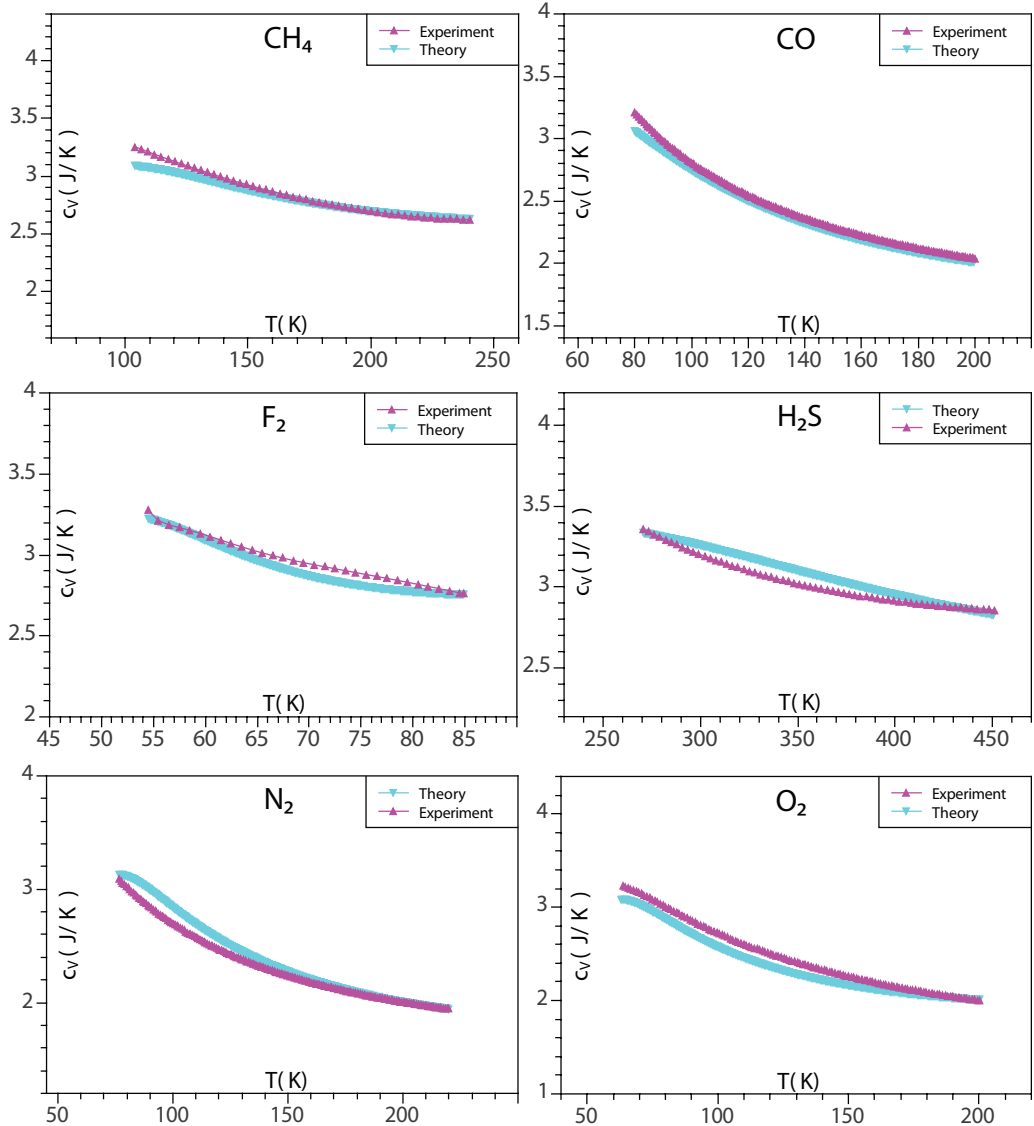}}}
\end{center}
\caption{Experimental and calculated $c_v$ for molecular liquids. Experimental $c_v$ and $\eta$ are taken from the NIST database \cite{nist} at pressures 50 MPa (CH$_4$), 55 MPa (CO), 0.1 MPa (F$_2$), and 170 MPa (H$_2$S), 65 MPa (N$_2$) and 45 MPa (O$_2$). Experimental $\eta$ for F$_2$ is from Ref. \cite{f2eta}. Values of $\tau_{\rm D}$ used in the calculation are 0.44 ps (CH$_4$), 0.55 ps (CO), 0.71 ps (F$_2$), 0.65 ps (H$_2$S), 0.38 ps (N$_2$) and 0.46 ps (O$_2$). Values of $G_{\infty}$ are 0.1 GPa (CH$_4$), 0.11 GPa (CO), 0.23 GPa (F$_2$), 0.15 GPa (H$_2$S), 0.14 GPa (N$_2$) and 0.22 GPa (O$_2$). Experimental values of $\alpha$ calculated from the NIST database at the corresponding pressures above are $3.1\cdot 10^{-3}$ K$^{-1}$ (CH$_4$), $4.1\cdot 10^{-3}$ K$^{-1}$ (CO), $4.31\cdot 10^{-3}$ K$^{-1}$ (F$_2$), $1.2\cdot 10^{-3}$ K$^{-1}$ (H$_2$S), $3.9\cdot 10^{-3}$ K$^{-1}$ (N$_2$) and $4.4\cdot 10^{-3}$ K$^{-1}$ (O$_2$). Values of $\alpha$ used in the calculation are $1.4\cdot 10^{-3}$ K$^{-1}$ (CH$_4$), $5.5\cdot 10^{-3}$ K$^{-1}$ (CO), $4.31\cdot 10^{-3}$ K$^{-1}$ (F$_2$), $6.5\cdot 10^{-4}$ K$^{-1}$ (H$_2$S), $3.9\cdot 10^{-3}$ K$^{-1}$ (N$_2$) and $4.4\cdot 10^{-3}$ K$^{-1}$ (O$_2$).
}
\end{figure*}

\begin{figure}
\begin{center}
{\scalebox{0.35}{\includegraphics{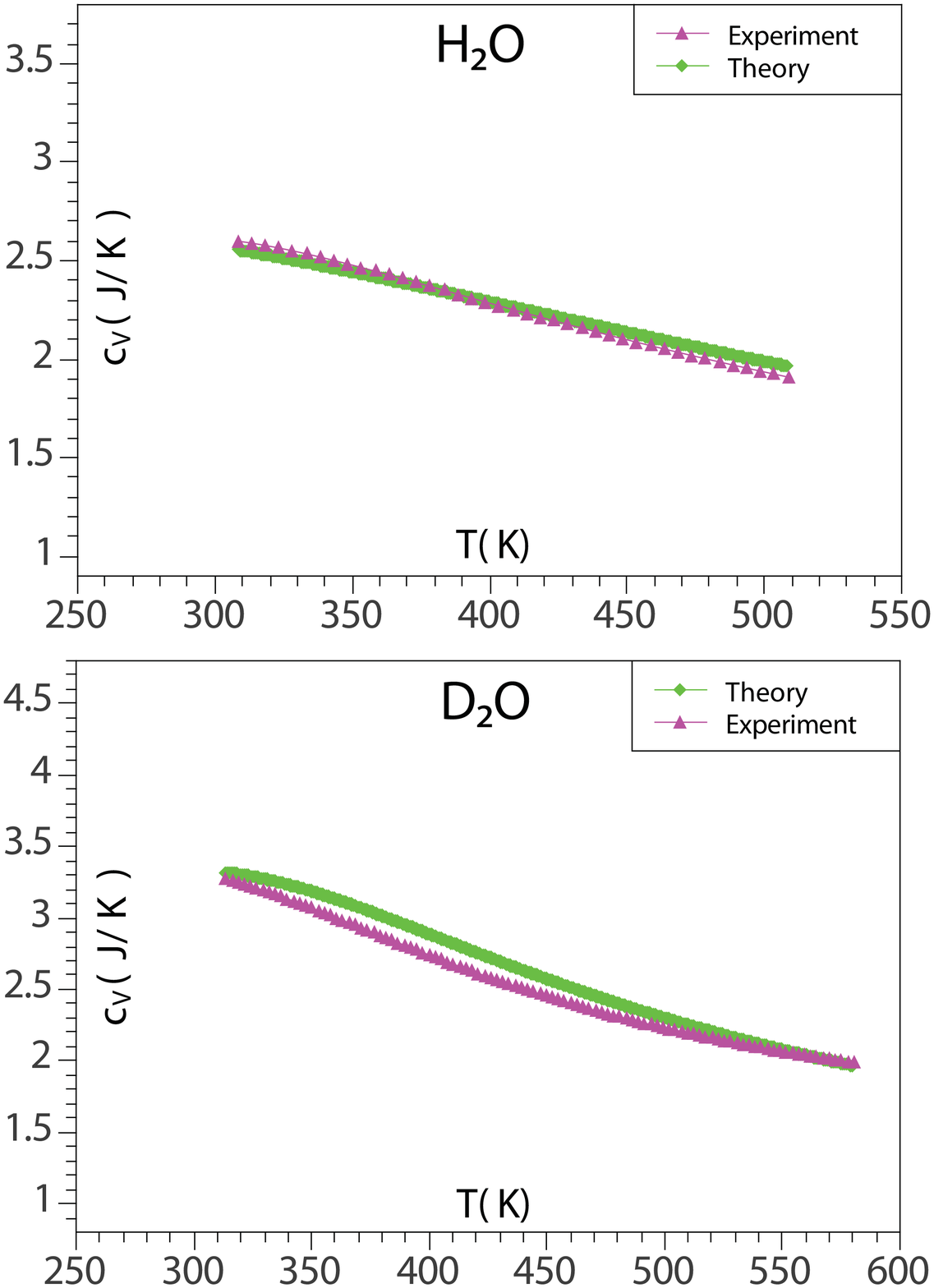}}}
\end{center}
\caption{Experimental and calculated $c_v$ for H$_2$O and D$_2$O. Experimental of $c_v$ and $\eta$ are taken from the NIST database \cite{nist} at pressures 150 MPa (H$_2$O) and 80 MPa (D$_2$O). Values of $\tau_{\rm D}$ used in the calculation are 0.15 ps (H$_2$O) and 0.53 ps (D$_2$O). For D$_2$O, experimental $\alpha$ calculated from the NIST database at the corresponding pressure above is $9.4\cdot 10^{-4}$ K$^{-1}$, and $\alpha$ used in the calculation is $1.1\cdot 10^{-3}$ K$^{-1}$. For H$_2$O, calculating $c_v$ in harmonic approximation ($\alpha=0$) gives good fit to experimental $c_v$.
}
\end{figure}

To calculate $c_v$ from Eq. (\ref{enf}), we have taken viscosity data from the NIST database at the same pressures as $c_v$ and converted it to $\tau$ using the Maxwell relationship $\tau=\frac{\eta}{G_{\infty}}$, where $\eta$ is viscosity and $G_{\infty}$ is infinite-frequency shear modulus, giving $\omega_{\rm F}=\frac{2\pi}{\tau}=\frac{2\pi G_{\infty}}{\eta}$. Viscosities of liquid metals and F$_2$ were taken from Refs. \cite{meta1,meta2,meta3,meta4,meta5,meta6} and Ref. \cite{f2eta}, respectively.

We note that Debye model is not a good approximation in molecular and hydrogen-bonded systems where the frequency of intra-molecular vibrations considerably exceeds the rest of frequencies in the system (e.g. 2260 K in O$_2$, 3340 K in N$_2$, 3572 K in CO). However, the intra-molecular modes are not excited in the temperature range of experimental $c_v$ (see Figures 3--4). Therefore, the contribution of intra-molecular motion to $c_v$ is purely rotational, $c^{rot}$. Being classical, the rotational motion gives $c^{rot}=R$ for linear molecules such as CO, F$_2$, N$_2$ and O$_2$ and $c^{rot}=\frac{3R}{2}$ for molecules with three rotation axes such as CH$_4$, H$_2$S, H$_2$O and D$_2$O. Consequently, $c_v$ for molecular liquids shown in Figures 3--4 correspond to heat capacities per molecule, with $c^{rot}$ subtracted from the experimental data. In this case, $N$ in Eq. (\ref{enf}) refers to the number of molecules.

We also note that in H$_2$O, approximately half of experimental $c_v$ is the configurational contribution due to the temperature-induced structural change of the hydrogen-bonded network \cite{kauzm}. The change includes changing coordinations of H$_2$O molecules during the continuous transition between the low-density and high-density liquid in the wide temperature range \cite{stanley}. Similarly to the phonon theory of solids, effects of this sort are not accounted for in the general theory presented here. Therefore, experimental $c_v$ for H$_2$O and D$_2$O shows $c_v$ with half subtracted from their values due to the configurational term and $c^{rot}=\frac{3R}{2}$ subtracted further as discussed above. As for other molecular liquids above, the resulting $c_v$ represents heat capacity per molecule. Due to the approximate way in which the configurational contribution is treated for H$_2$O and D$_2$O, the agreement between the experimental and calculated c$_v$ should be viewed as qualitative, in contrast to the quantitative agreement for the rest of liquids considered.

Experimental and calculated $c_v$ for 21 noble, metallic, molecular and hydrogen-bonded network liquids are shown in Figures 1--4. Figures 1--4 make one of the central points of our paper: the proposed phonon theory of liquids gives good agreement with experimental data.

Importantly, calculated $c_v$ has no free adjustable parameters, but depends on $\omega_{\rm D}$, $\alpha$ and $G_{\infty}$ (see Eq. (\ref{enf})) which are fixed by system properties. $\omega_{\rm D}$, $\alpha$ and $G_{\infty}$ that give the best fits in Figures 1--4 are in good agreement with their typical values. $\tau_{\rm D}=\frac{2\pi}{\omega_{\rm D}}$ used in the calculation (see the captions in Figures 1--4) are consistent with the known values that are typically in the 0.1--1 ps range \cite{g2,pilgrim}. For monatomic liquids, $\tau_{\rm D}$ were taken as experimental $\tau_{\rm D}$ in corresponding solids. Similarly, the difference between experimental $\alpha$ and $\alpha$ used in the calculation is within 30\% on average. Finally, $G_{\infty}$ used in the calculation are on the order of GPa typically measured \cite{g2,pilgrim}.

\section{Discussion}

We first note that some of the considered liquids are in the quantum regime at low temperature. Taking $T$ as the lowest temperature in Figures 1--4 and $\omega_{\rm D}=\frac{2\pi}{\tau_{\rm D}}$, where $\tau_{\rm D}$ are given in the caption of Figure 1--4, we find that $\frac{\hbar\omega_{\rm D}}{T}$ for various liquids varies in the range 0.1--3. Consequently, some liquids can be described by classical approximation fairly well, whereas others can not (for example, $\frac{\hbar\omega_{\rm D}}{T}>1$ for Ne, O$_2$, N$_2$, $F_2$, CH$_4$ and CO), implying progressive phonon excitation with temperature increase.

We observe that $c_v$ decreases with temperature in Figures 1--4. There are two main competing effects that contribute to $c_v$ in Eq. (\ref{enf}). First, temperature increase results in the increase of $\omega_{\rm F}=\frac{2\pi}{\tau}$. Consequently, $c_v$ decreases as a result of the decreasing number of shear waves that contribute to liquid energy and $c_v$. Second, $c_v$ increases with temperature due to progressive phonon excitation as discussed above. The first effect dominates in the considered temperature range, and the net effect is the decrease of $c_v$ with temperature seen in Figure 1. We further observe that $c_v$ changes from approximately 3 to 2, showing a universal trend across a wide range of liquids. This can be understood by noting that at low temperature, $\left(\frac{\omega_{\rm F}}{\omega_{\rm D}}\right)^3$ and the second term in the last bracket in Eq. (\ref{enf}) are small, giving $c_v$ close to 3 that includes the contribution from the anharmonic term. At high temperature when $\omega_{\rm F}\approx\omega_{\rm D}$ and Debye functions are close to 1, the term in the last bracket in Eq. (\ref{enf}) is 2, giving $c_v\approx 2$.

We note that on further temperature increase, $c_v$ decreases from 2 to its gas value of 1.5 \cite{nist}. This is related to the progressive disappearance of longitudinal phonons, an effect discussed in our forthcoming paper.

The decrease of $c_v$ seen in Figures 1--4 does not need to be generic. Indeed, when the phonon excitation dominates at low temperature, as in the case of strongly quantum liquids such as H$_2$ or He, $c_v$ increases with temperature \cite{nist}.

Good agreement between theoretical and calculated $c_v$ makes several important points. First, it is interesting to revisit the statement that liquid energy can not be calculated in general form because interactions are both strong and system-specific \cite{landau}. In the proposed phonon theory of liquids, this difficulty does not arise because strong interactions are treated from the outset, as in the theory of solids.

Second, the proposed phonon theory of liquids has an advantage over the previous approach where liquid potential energy is calculated from correlation functions and interatomic interactions \cite{landau,march}. Starting from the earlier proposals \cite{born}, this approach was developed in several directions (see, e.g., Refs. \cite{landau,march,leb,ash,rosen}), and can be used to calculate the energy of simple liquids where interactions are pair and short-ranged and correlations are two-body as is the case in, for example, noble liquids or hard-sphere models. The calculations become intractable in general case \cite{march,rosen}, particularly when interactions and correlations become complex (e.g., many-body) as in the liquids discussed above, precluding the calculation of $c_v$ in this approach. On the other hand, if the phonon states of the liquid depend on $\tau$ only as proposed by Frenkel \cite{frenkel}, the liquid energy and $c_v$ depend implicitly on $\tau$ only, even though correlation functions and interatomic interactions may affect both $c_v$ and $\tau$ in a complex way. Then, the relationship between $c_v$ and $\tau$ becomes fairly simple (see Eq. (\ref{enf})) and explains the experimental behavior of a wide class of liquids, both simple and complex.

The last assertion is important for a general outlook at liquids: despite their apparent complexity, understanding their thermodynamics may be easier than previously thought. Indeed, we have good understanding of thermodynamics and $c_v$ in solids based on phonons no matter how complicated interactions or structural correlations in a solid are. Our current results suggest that the same can apply to liquids. 

We are grateful to A. A. Abrikosov, M. T. Dove, A. Michaelides and J. C. Phillips for discussions.

\end{document}